Title: Cross Sections from 800 MeV Proton Irradiation of Terbium

Author(s):
Engle, Jonathan
Mashnik, Stepan G.
Bach, Hong
Couture, Aaron J.
Jackman, Kevin R.
Gritzo, Russell E.
Ballard, Beau
Fassbender, Michael E.
Smith, Donna M.
Bitteker, Leo J. Jr.
Ullman, Kenneth J.
Gulley, Mark S.
Pillai, Chandra
John, Kevin D.
Birnbaum, Eva R.
Nortier, Francois M.

# Cross Sections from 800 MeV Proton Irradiation of Terbium

J.W. Engle*, S.G. Mashnik, H. Bach, A. Couture, K. Jackman, R. Gritzo, B. D. Ballard, M. Faßbender, D.M. Smith, L.J. Bitteker, J.L. Ullmann, M. Gulley, C. Pillai, K. D. John, E. R. Birnbaum, F.M. Nortier**

*Los Alamos National Laboratory, Los Alamos, New Mexico, 87545, USA*

## Abstract

A single terbium foil was irradiated with 800 MeV protons to ascertain the potential for production of lanthanide isotopes of interest in medical, astrophysical, and basic science research and to contribute to nuclear data repositories. Isotopes produced in the foil were quantified by gamma spectroscopy. Cross sections for 36 isotopes produced in the irradiation are reported and compared with predictions by the MCNP6 transport code using the CEM03.03, Bertini and INCL+ABLA event generators. Our results indicate the need to accurately consider fission and fragmentation of relatively light target nuclei like terbium in the modeling of nuclear reactions at 800 MeV. The predictive power of the code was found to be different for each event generator tested but was satisfactory for most of the product yields in the mass region where spallation reactions dominate. However, none of the event generators' results are in complete agreement with measured data.

* Corresponding author.
*E-mail address*: jwengle@lanl.gov.
** Principle investigator.
*E-mail address*: meiring@lanl.gov.

## 1. Introduction

The 800 MeV proton beam at the Los Alamos Neutron Science Center (LANSCE) facility has been used to make a wide variety of radionuclides since the early 1970s [1]. This manner of isotope production utilizes spallation reactions, which occur in two main stages. The intranuclear cascade involves incident particles interacting with individual nucleons, instead of the nucleus as a whole. Several high-energy particles can leave the nucleus and potentially initiate further spallation reactions in neighboring nuclei, resulting in a chain reaction process that eventually dies out when secondary particles no longer have sufficient energy to initiate a spallation event. The nucleus involved in the spallation reaction is left in an excited state and relieves its excitation energy by competing processes of evaporation or fission. If the excitation energy of the residual nucleus produced after the intranuclear cascade stage of a reaction is of the order of tens of MeV or greater, preequilibrium emission of particles is also possible during the equilibration of the nucleus, before evaporation of particles or fission of the compound nucleus.

Cross sections for proton-induced radionuclide production at 800 MeV and nearby energies have been measured at several laboratories using different methods, depending on the target-nuclei and the measured isotopes. Most of the residual nuclides from proton-induced reactions were quantified by x-ray and γ-spectroscopy (see, e.g., [1-3] and references therein). A few measurements are reported using chemical treatment of the irradiated samples, or/and γ-spectroscopy followed by accelerator mass spectrometry (see, e.g., [4] and references therein). Nuclide production cross sections from $^{197}$Au at 800 MeV and from $^{56}$Fe at 750 MeV were obtained by the inverse kinematics method at GSI in Darmstadt, Germany [5]. Production of the alpha emitter $^{148}$Gd at 800 and 600 MeV from several target-nuclei was measured using charged-particle spectroscopy at Los Alamos [6].

Unfortunately, reported measurements of experimental cross sections for nuclide production from terbium are very scarce. We are aware of only two studies that measured cumulative or independent cross sections for proton reactions with terbium [7, 8]: the formation of $^{83,84,86}$Rb was measured for proton energies between 0.6 and 21 GeV at CERN [7] and the $^{159}$Tb(p,x)$^{149g}$Tb cross section was measured by Mironov, et al. at the St. Petersburg Institute of Nuclear Physics (Gatchina) in Russia from 60 MeV to 1 GeV [8]. Finally, the proton-induced fission cross section of Tb was measured at 600 MeV (1.9 ± 0.2 mb) [9] and at 1 GeV (9.0 ± 1.5 mb) [10].

This work is an attempt to reduce gaps in reported literature by measuring cross sections of products produced by 800 MeV proton bombardment of a thin Tb target. The additional data reported here are intended to support efforts aimed at improving present theoretical models used to simulate spallation, fission and fragmentation of Tb and other nearby target-nuclei irradiated with charged particles. These data are also potentially useful to accelerator facilities targeting production-scale yields of several isotopes of interest.

We compare our measured cross sections with predictions of the MCNP6 transport code [11, 12] using three different event generators available in MCNP6 to simulate high energy nuclear reactions. All predictions were obtained prior to the measurement. A brief description of the three event generators follows:

1) The default MCNP6 option, which for our reaction is the Bertini IntraNuclear Cascade (INC) [13], followed by the Multistage Preequilibrium Model (MPM) [14], followed by the evaporation model as described with the EVAP code by Dresner [15], followed by or in competition with the RAL fission model [16] (if the charge of the compound nucleus Z is ≥ 70), referred to herein simply as "Bertini".
2) The improved Cascade-Exciton Model (CEM) of nuclear reactions as implemented in the code CEM03.03 [17, 18].
3) The Intra-Nuclear Cascade model developed at Liege (INCL) by Cugnon with coauthors [19] merged with the evaporation-fission model ABLA [20] developed at GSI, Darmstadt, Germany, referred to herein as "INCL+ABLA".

These event generators have previously been benchmarked against a large variety of experimental data and compared with each other and several other modern models (see, e.g., [3] and references therein). However, very few data were available in the past to test the models with reactions involving Tb nuclei; experimental data reported here attempt to address this deficiency.

## 2. Material and methods

### 2.1 Irradiation

Terbium target foils were irradiated with 800 MeV $^{1}H^{+}$ ions at the LANSCE accelerator facility in the Weapons Neutron Research (WNR) Blue Room. Terbium discs with a diameter of 24.7 ± 0.1 mm, a thickness of 80.4 ± 0.1 mg/cm$^2$ and a purity of 99.9% were obtained from American Elements (Los Angeles, CA). High purity (99.999%) aluminum monitor foils with the same diameter and a thickness of 63.5 ± 0.1 mg/cm$^2$ were obtained from the same supplier and used to verify facility-reported proton fluence. Target foils were sandwiched between single 25 μm thick layers of Kapton tape, which served as a catcher for recoil ions. Two target foil stacks were irradiated in the measurement of the data reported here. The first stack included only two terbium foils and accepted a beam current of 49 ± 3 nA for approximately 0.5 hours. The second stack, comprised of 10 terbium foils, was irradiated for 10.4 ± 0.1 hours with an average beam current of 50 ± 4 nA; the purpose of this second irradiation was to produce larger amounts of longer-lived radioisotopes for quantification after a long period of decay. Both stacks included empty target holders as sample blanks, which were counted in the same manner as the terbium samples to quantify the incidental presence or production of radioisotopes (e.g., $^{7}Be$) from background sources or from nucleon interactions with the Kapton, adhesive, or aluminum holders.

Prior to the experiment, a stainless steel (SS) monitor foil was activated in the beam to establish the beam position in the target irradiation station by exposing gafchromic film to the irradiated foils. Additional SS foils were included in the terbium foil stack to allow verification of the beam position on the terbium foils in same manner.

Using accelerator parameters and the stopping power formalism published by Anderson and Ziegler [21], the incident proton energy on the first foil was calculated to be 795.4 ± 0.5 MeV. For convenience, a nominal value of 800 MeV is used when describing this work (the exact 800 MeV value is used in all MCNP6 simulations).

The established proton beam monitor reaction $^{27}Al(p,x)^{22}Na$ [22] was used to quantify the integrated proton flux, employing a value of 15.0 ± 0.9 mb based on a previous evaluation of the literature data [23].

## 2.2 Gamma spectroscopy

After irradiation, the activation products were transported to the Nuclear and Radiochemistry Group (C-NR) Countroom, where they were repeatedly assayed by non-destructive gamma spectroscopy for more than 2 months. The HPGe detector used to assay the foils is a p-type Al-window ORTEC GEM detector with a relative efficiency at 1333 keV of about 10% and a measured FWHM at 1333 keV of 1.99 keV. Spectra collection times varied from 20 to 120 minutes for foils from the first stack and 1000-2000 minutes for foils from the second, thicker stack, and source-to-detector distances varied based on detector dead times. Changes in spectra backgrounds, detector resolution, and energy calibration (gain), were checked daily. Detector efficiency was calibrated prior to the beginning of data collection and verified after the experiment's completion.

An in-house developed analysis code was used to process the resulting spectra. GAMANAL is a sophisticated spectroscopy analysis code that has been in use since the early 1970s [24]. In the 1980s this software package was modified to handle large volumes of data in the C-NR Countroom. The revised code, SPECANAL, was used to extract photopeak areas from gamma spectra for this work. SPECANL calculates gamma ray spectral peaks' integrated counts by fitting data with modified Gaussian functions with low-energy exponential tails. Photopeak backgrounds are approximated using a photopeak-width broadened linear step function. This background is subtracted and values are corrected for detector deadtime and photopeak efficiency. Linear regressions were applied to the integrated area of each photopeak logged over the duration of the counting period. Gamma energies and intensities used in this work were taken from the National Nuclear Data Center's (NNDC) online archives and are listed in Table 1 [25]. The activity at the end of bombardment (EoB) of each isotope of interest was determined by fitting of its decay curve and cross sections were calculated using the well-known activation formula.

Uncertainties in linear regressions' fitted parameters were computed from covariance matrices as the standard error in the activity extrapolated to the end of bombardment. This value was combined according to the Gaussian law of error propagation with estimated uncertainties from detector calibration and geometry (2.8% combined), foil dimensions (0.1%), and proton flux (6.7% and 8.1% for short and long irradiations, respectively). Multiple photopeaks were used (up to a maximum of 4) when possible, so additional error as the standard deviation of these complimentary measurements was combined with the errors described above, again according to the Gaussian law of error propagation.

## 2.3 High Energy Transport Code Predictions

As mentioned in the introduction, the measured results are compared with the predictions of the latest Los Alamos Monte Carlo transport code MCNP6 [11, 12] using three different event generators available in MCNP6. All the models employed by the event generators have seen decades of widespread use in both applied and academic studies and are described in detail elsewhere (see, e.g. [13-20] and references therein).

The improved Cascade-Exciton Model (CEM) as implemented in the code CEM03.03 [17, 18] calculates nuclear reactions induced by nucleons, pions, and photons. It assumes that the reactions occur generally in three stages. The first stage is the INC, in which primary particles can be re-scattered and produce secondary particles several times prior to absorption by (or escape from) the nucleus. When the cascade stage of a reaction is completed, CEM03.03 uses the coalescence model to "create" high-energy d, t, $^{3}$He, and $^{4}$He by final-state interactions among emitted cascade nucleons. The emission of the cascade particles determines the particle-hole configuration, Z, A, and the excitation energy that is the starting point for the second, preequilibrium stage of the reaction. The subsequent relaxation of the nuclear excitation is treated with an improved version of the modified exciton model of preequilibrium decay followed by the equilibrium evaporation/fission stage (also called the compound nucleus stage), which is described with an extension of the Generalized Evaporation Models (GEM) code, GEM2, by Furihata [26]. Generally, all four components may contribute to experimentally measurable particle spectra and other distributions. But if the residual nuclei after the INC have atomic numbers in the range $A < 13$, CEM03.03 uses the Fermi breakup model [27] to calculate their further disintegration instead of using the preequilibrium and evaporation models. Fermi breakup is faster to calculate than GEM and gives results similar to the continuation of the more detailed models to much lighter nuclei.

To test the implementation of CEM03.03 in MCNP6, as a Validation and Verification (V&V) exercise for MCNP6, we have performed the same calculations with MCNP6 using the CEM03.03 event generator, as well as with CEM03.03 used as a separate stand-alone code [17, 18], outside of MCNP6. The results obtained in these two different ways are nearly identical; below, for clarity, we present only results calculated with MCNP6 using CEM03.03.

The MCNP6 default option for proton-induced reactions at energies below 3.5 GeV is the Bertini INC [13]. By default, Bertini INC is followed by the Multistage Preequilibrium Model (MPM) [14]. The relaxation of an excited compound nucleus produced after the preequilibrium stage of a reaction is calculated with the Weisskopf evaporation model as implemented in the EVAP code by Dresner [15]. If the charge of the compound nucleus $Z \geq 70$, then a competition between evaporation and fission is taken into account, with the latter calculated using the RAL fission model by Atchison [16]. This is not the case for our reaction because even absorption of the incident 800 MeV proton by the terbium nucleus without emission of protons or complex particles would require emission of four negative pions in order to form a nucleus with $Z = 70$. For this reason the Bertini option of MCNP6 is not expected to account for fission and therefore does not predict fission fragments from our reaction. The Bertini event generator option also accounts for Fermi breakup of excited nuclei when $A < 18$; it uses a realization of the Fermi breakup model as described in [28]. However, the Bertini option of MCNP6 does not account for the coalescence of complex particles.

The version of the Intra-Nuclear Cascade model developed at Liege (INCL) by Joseph Cugnon with coauthors [19] is usually used to describe reactions induced by nucleons and complex-particles up to $^{4}$He at incident energies up to several GeV. In MCNP6, it is merged with the evaporation-fission model ABLA [20] developed at GSI in Darmstadt,

Germany. The version of INCL+ABLA used currently in MCNP6 accounts for possible fission of compound nuclei produced in our reaction, but it does not account for preequilibrium processes, for Fermi break-up of light residual nuclei, or for coalescence of complex particles after INC. A future realization of MCNP6 will incorporate an improved version of INCL+ABLA (see [11, 12]).

All event generators used compute only independent cross sections; cumulative cross sections were subsequently calculated using these independent values summed separately according to the decay behavior of parent products using the Table of Isotopes [29].

## 3. Results

Representative fits of photopeaks' integrated area over time for four isotopes are included below in Fig. 1.

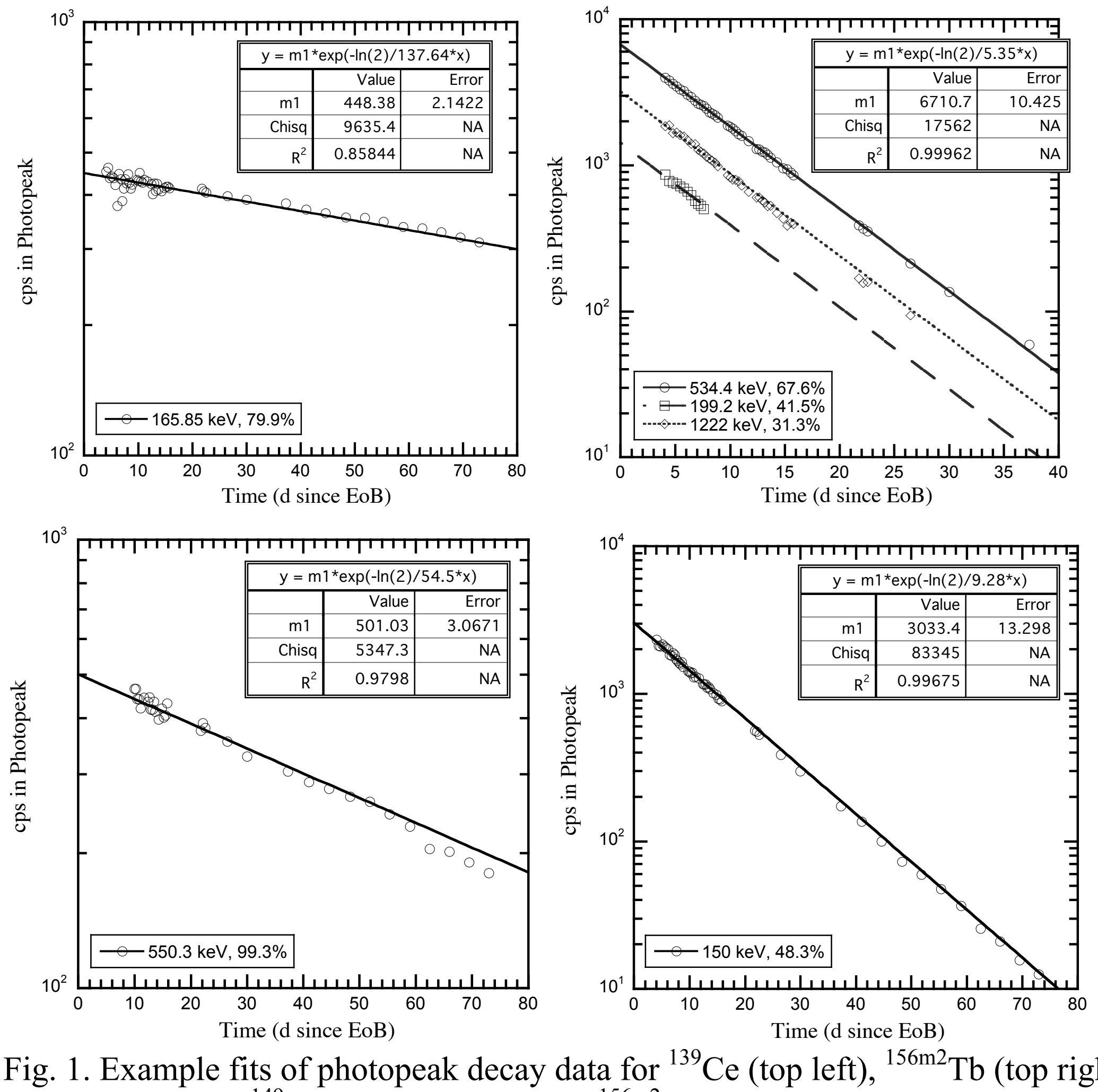

Fig. 1. Example fits of photopeak decay data for $^{139}$Ce (top left), $^{156m2}$Tb (top right), $^{148}$Eu (bottom left), and $^{149}$Gd (bottom right). For $^{156m2}$Tb, only the fitting parameters for the 534.4 keV peak are included.

Measured cross sections are reported in Table 1. Most of the measured cross sections are cumulative. Table 1 presents cross sections from CEM03.03, Bertini, and INCL+ABLA models calculated within MCNP6. All calculations used the models' default parameters. Fig. 2 compares product yield mass distributions predicted by CEM03.03, Bertini, and INCL+ABLA with our measured cumulative cross sections and includes a single point in the fragmentation region for $^{7}$Be. The $^{159}$Tb(p,x)$^{83}$Rb experimental value at 600 MeV was taken from Ref. [7].

Table 1. A comparison of measured and calculated cross sections from CEM03.03, Bertini+Dresner+MPM+RAL, and INCL+ABLA event generators of MCNP6 for 800 MeV proton irradiation of a thin terbium foil.

| | | Measured cross sections | | | MCNP6 calculated cross sections | | | | | |
|---|---|---|---|---|---|---|---|---|---|---|
| | | | | | CEM03.03 | | Bertini | | INCL+ABLA | |
| Isotope | Type [a)] | σ (mb) | Δσ (mb) | γ (keV) [b)] and $I_\gamma$ (%) | σ (mb) | Δσ (mb) | σ (mb) | Δσ (mb) | σ (mb) | Δσ (mb) |
| $^{7}$Be | i | 8.0 | 0.5 | 478 (*10.4*) | 0.87 | 0.01 | - | - | 9.78e-4 | 4.89e-4 |
| $^{54}$Mn | i | 0.11 | 0.01 | 835 (*100*) | 0.035 | 0.002 | - | - | 2.2e-3 | 0.7e-3 |
| $^{60}$Co | c | 0.035 | 0.004 | 1173.2 (*99.9*) | 0.029 | 0.003 | - | - | 6.6e-3 | 2.0e-3 |
| | | | | 1332.5 (*100*) | | | | | | |
| $^{113}$Sn | c | 3.1 | 0.3 | 392 (*64.0*) | 6.85 | 0.54 | 4.74 | 0.52 | 0.17 | 0.04 |
| $^{127}$Xe | c | 26.2 | 2.4 | 172 (*68.7*) | 26.49 | 0.13 | 20.50 | 0.13 | 10.89 | 0.11 |
| | | | | 203 (*25.7*) | | | | | | |
| $^{129}$Cs | c | 38.7 | 2.9 | 372 (*30.6*) | 30.06 | 0.15 | 22.42 | 0.14 | 15.72 | 0.13 |
| | | | | 411 (*22.3*) | | | | | | |
| $^{128}$Ba | c | 29.1 | 1.9 | 273 (*14.5*) | 30.45 | 0.14 | 23.72 | 0.13 | 12.28 | 0.10 |
| $^{131}$Ba | c | 36.7 | 3.9 | 133 (*2.2*) | 33.83 | 0.15 | 24.63 | 0.15 | 21.28 | 0.15 |
| | | | | 216 (*20.4*) | | | | | | |
| | | | | 496 (*48.0*) | | | | | | |
| $^{133}$Ba | c | 73.3 | 53.6 | 356 (*64.1*) | 35.26 | 0.14 | 26.92 | 0.17 | 27.54 | 0.18 |
| $^{139}$Ce | c | 56.3 | 3.8 | 166 (*79.9*) | 40.81 | 0.15 | 44.85 | 0.19 | 39.68 | 0.23 |
| $^{141}$Ce | c | 2.3 | 0.3 | 145 (*48.3*) | 0.025 | 0.002 | 0.132 | 0.003 | 0.023 | 0.007 |
| $^{143}$Pm | c | 64.7 | 9.4 | 742 (*38.5*) | 39.33 | 0.15 | 39.28 | 0.18 | 48.07 | 0.22 |
| $^{144}$Pm | c | 7.2 | 0.9 | 618 (*98.3*) | 3.05 | 0.02 | 4.61 | 0.03 | 7.19 | 0.04 |
| | | | | 696 (*99.5*) | | | | | | |
| $^{146}$Pm | i | 0.51 | 0.03 | 454 (*65.0*) | 1.01 | 0.01 | 1.64 | 0.02 | 3.60 | 0.03 |
| $^{149}$Pm | c | 14.8 | 10.0 | 286 (*3.1*) | 0.23 | 0.07 | 0.52 | 0.14 | 1.19 | 0.22 |
| $^{145}$Sm | c | 66.7 | 47.8 | 61 (*12.2*) | 38.27 | 0.14 | 36.96 | 0.16 | 45.96 | 0.20 |
| $^{153}$Sm | c | 7.0 | 0.6 | 103 (*29.3*) | 0.16 | 0.01 | 0.80 | 0.02 | 1.72 | 0.03 |
| $^{145}$Eu | c | 39.5 | 3.4 | 654 (*15.1*) | 31.46 | 0.11 | 27.21 | 0.12 | 31.47 | 0.14 |
| | | | | 894 (*66.4*) | | | | | | |
| | | | | 1659 (*14.9*) | | | | | | |
| $^{146}$Eu | c | 22.5 | 3.0 | 634 (*80.9*) | 35.68 | 0.12 | 31.96 | 0.13 | 36.26 | 0.16 |
| | | | | 747 (*99.3*) | | | | | | |
| $^{147}$Eu | c | 57.9 | 4.6 | 197 (*24.4*) | 33.09 | 0.13 | 32.23 | 0.15 | 39.83 | 0.17 |
| | | | | 677 (*9.0*) | | | | | | |
| $^{148}$Eu | c | 19.6 | 1.4 | 414 (*20.4*) | 7.55 | 0.03 | 10.56 | 0.05 | 17.03 | 0.06 |
| | | | | 550 (*99.3*) | | | | | | |
| | | | | 630 (*71.9*) | | | | | | |
| $^{149}$Eu | c | 66.4 | 6.0 | 327 (*4.0*) | 35.30 | 0.14 | 35.41 | 0.18 | 44.54 | 0.20 |
| $^{150}$Eu | i | 2.3 | 0.2 | 439.4 (*80.3*) | 5.23 | 0.03 | 7.41 | 0.04 | 13.29 | 0.06 |
| $^{155}$Eu | c | 20.7 | 4.4 | 87 (*30.7*) | 0.84 | 0.01 | 4.66 | 0.04 | 6.57 | 0.05 |
| $^{156}$Eu | c | 4.0 | 1.8 | 646 (*6.3*) | 0.21 | 0.01 | 2.79 | 0.03 | 4.72 | 0.04 |
| | | | | 812 (*9.7*) | | | | | | |
| | | | | 1153.7 (*6.8*) / 1154.1 (*4.7*) | | | | | | |
| $^{146}$Gd | c | 26.0 | 2.8 | 115 (*88.0)* | 24.40 | 0.08 | 18.24 | 0.08 | 16.97 | 0.09 |
| | | | | 155 (*45.1)* | | | | | | |
| $^{147}$Gd | c | 31.2 | 2.8 | 229 (*58.4*) | 21.89 | 0.09 | 17.45 | 0.09 | 21.46 | 0.11 |
| | | | | 396 (*31.4*) | | | | | | |
| | | | | 929 (*18.4*) | | | | | | |
| $^{149}$Gd | c | 42.6 | 3.8 | 150 (*48.3*) | 27.60 | 0.11 | 24.52 | 0.12 | 29.35 | 0.13 |
| | | | | 298 (*28.6*) | | | | | | |
| | | | | 347 (*23.9*) | | | | | | |
| $^{151}$Gd | c | 46.7 | 5.1 | 123 (*5.6*) | 30.40 | 0.12 | 28.62 | 0.14 | 35.28 | 0.15 |
| $^{153}$Gd | c | 65.6 | 7.5 | 97 (*29.0*) | 34.70 | 0.13 | 32.24 | 0.15 | 39.76 | 0.17 |
| | | | | 103 (*21.1*) | | | | | | |
| $^{151}$Tb | c | 32.0 | 15.8 | 252 (*26.3*) | 18.30 | 0.08 | 15.23 | 0.08 | 15.51 | 0.08 |
| | | | | 287 (*28.3*) | | | | | | |
| | | | | 479 (*15.4*) | | | | | | |
| $^{152}$Tb | i(g) | 29.7 | 7.7 | 160 (*16.5*) | 13.22 [c)] | 0.04 [c)] | 12.29 [c)] | 0.06 [c)] | 15.05 [c)] | 0.06 [c)] |
| | | | | 283 (*59.7*) | | | | | | |
| $^{153}$Tb | c | 44.4 | 21.6 | 170 (*6.3*) | 22.77 | 0.09 | 20.12 | 0.10 | 20.02 | 0.10 |
| | | | | 534 (*41.5*) | | | | | | |
| $^{155}$Tb | c | 42.0 | 4.8 | 85 (*32.0*) | 37.01 | 0.10 | 26.08 | 0.11 | 25.76 | 0.11 |
| | | | | 105 (*25.1*) | | | | | | |
| $^{156}$Tb | i(m2) | 34.4 | 10.6 | 199 (*41.5*) | 16.40 [d)] | 0.05 [d)] | 20.48 [d)] | 0.07 [d)] | 27.28 [d)] | 0.08 [d)] |
| | | | | 534 (*67.6*) | | | | | | |
| | | | | 1222 (*31.3*) | | | | | | |
| $^{158}$Tb | i | 0.53 | 0.04 | 944.2 (*43.9*) | 43.14 | 0.08 | 71.32 | 0.13 | 72.74 | 0.13 |

[a)] Refers to the type of cross section measured: i = independent; c = cumulative.
[b)] Refers to gamma rays used for quantification.
[c), d)] All models provide only the sum of g, m2, and other possible final states of products; i.e., the $^{152}$Tb and $^{156}$Tb theoretical yields may be overestimating the measured g and m2 cross sections, respectively.

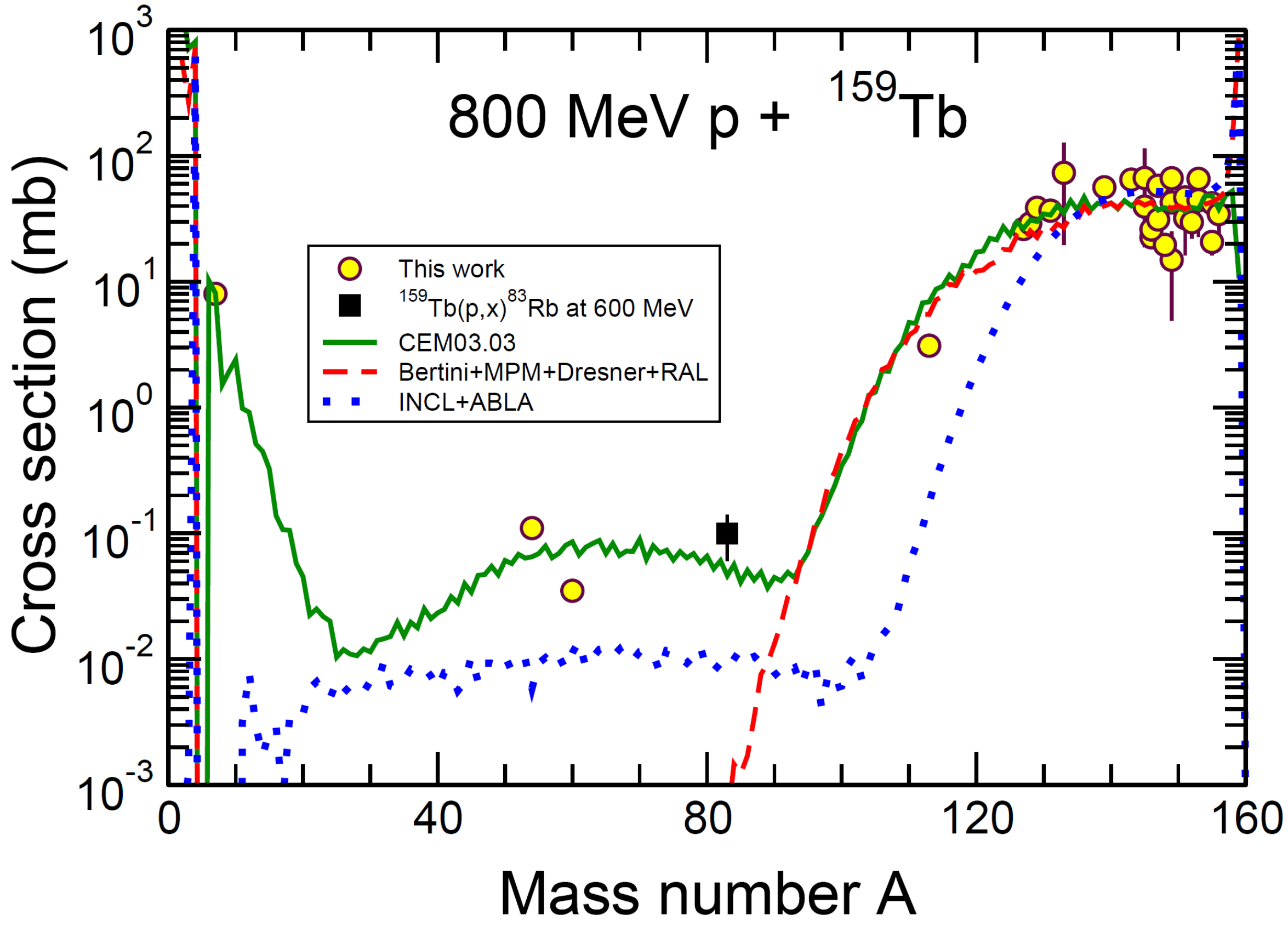

Fig. 2. Comparison of mass distributions of product yields predicted by CEM03.03, Bertini+MPM+Dresner+RAL and INCL+ABLA from 800 MeV p + $^{159}$Tb with cumulative cross sections measured in the present work. The only product yield in the fragmentation region measured here, the independent cross section of $^{7}$Be, and the independent cross section of $^{54}$Mn produced via fission, are shown as well, for comparison. The $^{159}$Tb(p,x)$^{83}$Rb data point at 600 MeV was measured by Lagarde-Simonoff and Simonoff and was published in Ref. [7].

## 4. Discussion

For all nuclides identified, this work presents the first reported measurement of cross sections for production from 800 MeV proton irradiation of terbium. In particular, our measured cross section for $^{7}$Be facilitates comparison of measured data with theoretical predictions in the fragmentation region where such a test was previously impossible. Two products, $^{60}$Co and $^{54}$Mn, were measured in the region of fission fragments, where data are particularly scarce. New data in the mass region near the boundary between spallation and fission events (e.g., for $^{113}$Sn) are also especially useful because it is in these regions where the model predictions vary most rapidly.

The approximately three-day period between the end of irradiation and the beginning of our earliest counting experiments prevents quantification of products with $t_{1/2} < 6$ h. As the atomic number of produced radionuclides becomes distant from 159, quantification similarly becomes increasingly challenging, with the best statistics collected for nuclides fed by long parent decay chains. The decay of Tb radioisotopes is particularly complex, with multiple branching ratios detracting from cumulative cross sections.

In the absence of multiple decay modes that reduce parents' contribution to daughter activities, cross sections are generally higher for nuclei approaching the stability line. In a few cases, useful data were extracted from contaminated photopeaks and reported after quantitative accounting for interferences. For example, $^{153}$Sm (46.3 h, 100% $\beta^-$) possesses a single useful γ-ray (103.18 keV, $I_\gamma$ = 29.25%); $^{153}$Gd (240.4 d, 100% EC) also emits a 103.18 keV γ ($I_\gamma$ = 21.1%). During curve fitting, the contribution from $^{153}$Gd was fixed to parameters identified from its undisturbed 97.4 keV γ signal ($I_\gamma$ = 29.0%) and subtracted from the 103.18 keV signal, lending clarity to $^{153}$Gd's obfuscation of the weak signal from $^{153}$Sm. Notably, quantification of assumed $^{22}$Na ($t_{1/2}$ = 2.603 y, 1274.54 keV γ) was not possible due to signal interference from $^{154}$Eu ($t_{1/2}$ = 8.601 y, 1274.43 keV γ).

Quantification of α-emitters (e.g. $^{154}$Dy, $^{148}$Gd, $^{150}$Gd, and $^{146}$Sm) and x-ray emitters (e.g. $^{159}$Dy, $^{158}$Tb, and $^{157}$Tb) requires destructive assays; such a study was outside the scope of the present work.

A comparison of the new measured data with MCNP6 predictions using CEM03.03, Bertini, and INCL+ABLA event generators is informative. Fig. 2 reveals good agreement between measurement and all three codes' predictions for products in the spallation region, $130 \leq A \leq 159$. CEM03.03 and Bertini predictions are in very good agreement in the lower mass region of transition between spallation and fission processes, down to A ~ 90, but INCL+ABLA results in this transition region differ by an order of magnitude from the CEM03.03- and Bertini-generated values. The INCL+ABLA prediction also underestimates the measured yield of $^{113}$Sn by approximately a factor of 10.

Agreement between CEM03.03 and Bertini results ends at A ~ 90. The RAL fission model [17] neglects fission of nuclei with atomic numbers lower than 70, and the Bertini+MPM+Dresner+RAL event generator does not predict fission fragments from Tb (for products with mass number A < 80, the Bertini code predicts only the formation of $^{1}$H, d, t, $^{3}$He, and $^{4}$He, all produced via evaporation from compound nuclei).

The INCL+ABLA event generator considers fission of Tb at our energy, as does CEM03.03. Measured fission fragment yields for $^{54}$Mn and $^{60}$Co are in good agreement with CEM03.03, as are previously measured values for the $^{159}$Tb(p,x)$^{83}$Rb cross section at 600 MeV [7]. The INCL+ABLA prediction is approximately an order of magnitude lower. The Bertini code predicts only a tiny production of $^{83}$Rb via deep spallation, more than two orders of magnitude below the measured value and does not allow for production of $^{54}$Mn or $^{60}$Co.

A qualitative comparison of predictions by CEM03.03 and INCL+ABLA with the very few fission data measured in Refs. [9, 10] at 600 MeV and 1 GeV is also instructive. CEM03.03 predicts a fission cross section of 1.82 ± 0.01 mb for 800 MeV p + $^{159}$Tb. This agrees reasonably well with the value of 1.9 ± 0.2 mb measured in Ref. [9] at 600 MeV, but it is almost five times lower than the 9.0 ± 1.5 mb measured by Vaishnene et al. at 1 GeV [10]. From Fig. 2 we see that the fission cross section calculated for our reaction by INCL+ABLA is almost an order of magnitude lower than the values predicted by CEM03.03.

An important question for applications involving complex geometries and high statistics simulations is the computing time required by different event generators in order to achieve the needed accuracy. The computing time required by the Bertini INC is usually similar to the time required by the INC of CEM03.03. For example, Bertini INC and CEM03.03 would require the same time to simulate a reaction on a light nucleus, e.g., carbon, because both models account only for Fermi breakup in addition to INC for nuclei with $A < 13$. The situation is different for $^{159}$Tb: both the CEM03.03 and Bertini options of MCNP6 consider INC and preequilibrium processes but CEM accounts additionally for the possible evaporation of up to 66 different types of elementary particles and light fragments up to $^{28}$Mg ejected from excited compound nuclei, including a possible fission and subsequent evaporation from excited fission fragment. However, the Bertini option considers evaporation of only 6 types of particles (n, p, d, t, $^{3}$He, and $^{4}$He) from compound nuclei, and the RAL fission model used with Bertini INC does not account for fission for $Z < 70$. As a result, the computing time required by the Bertini option is much shorter than that needed for CEM03.03. We performed all the calculations for the present work ($10^7$ simulated inelastic "histories" for each event generator) in parallel, with MPI, on the "Turing" supercomputer available at LANL, using 4 nodes and 64 processors. The total computing time reported in the output files by CEM03.03, Bertini, and INCL+ABLA was 302.90, 32.52, and 681.94 minutes, respectively.

Fig. 3 compares measured cross sections with theoretical results for individual isotopes and shows general agreement (within a factor of two) for isotopes with mass near that of the target nucleus. CEM03.03 agrees better than other models with many measured cross sections in the spallation region but under-predicts by more than an order of magnitude the yields of $^{155}$Eu$^{c}$ and $^{165}$Eu$^{c}$. These isotopes are likely produced by peripheral interactions of the bombarding protons with Tb nuclei, involving (p,p$^{3}$He), (p,p$^{4}$He), and (p,d$^{4}$He) reactions in which the ejected particles are expected to carry large amounts of energy, leaving the residual nucleus with little excitation energy. The semi-classical INC approximation of CEM03.03 is clearly too rough to characterize such reactions accurately. All event generators, and especially CEM03.03, under-predict the cross sections of $^{141}$Ce$^{c}$, $^{149}$Pm$^{c}$, and $^{153}$Sm$^{c}$. Furthermore, CEM03.03 predicts a cross section for $^{7}$Be that underestimates the measured value by a factor of nine. INCL+ABLA predictions are several orders of magnitude lower still, and Bertini predicts no yield of $^{7}$Be. This more detailed analysis of results clearly demonstrates that models must be improved in order to accurately predict yields of isotopes from arbitrary reactions.

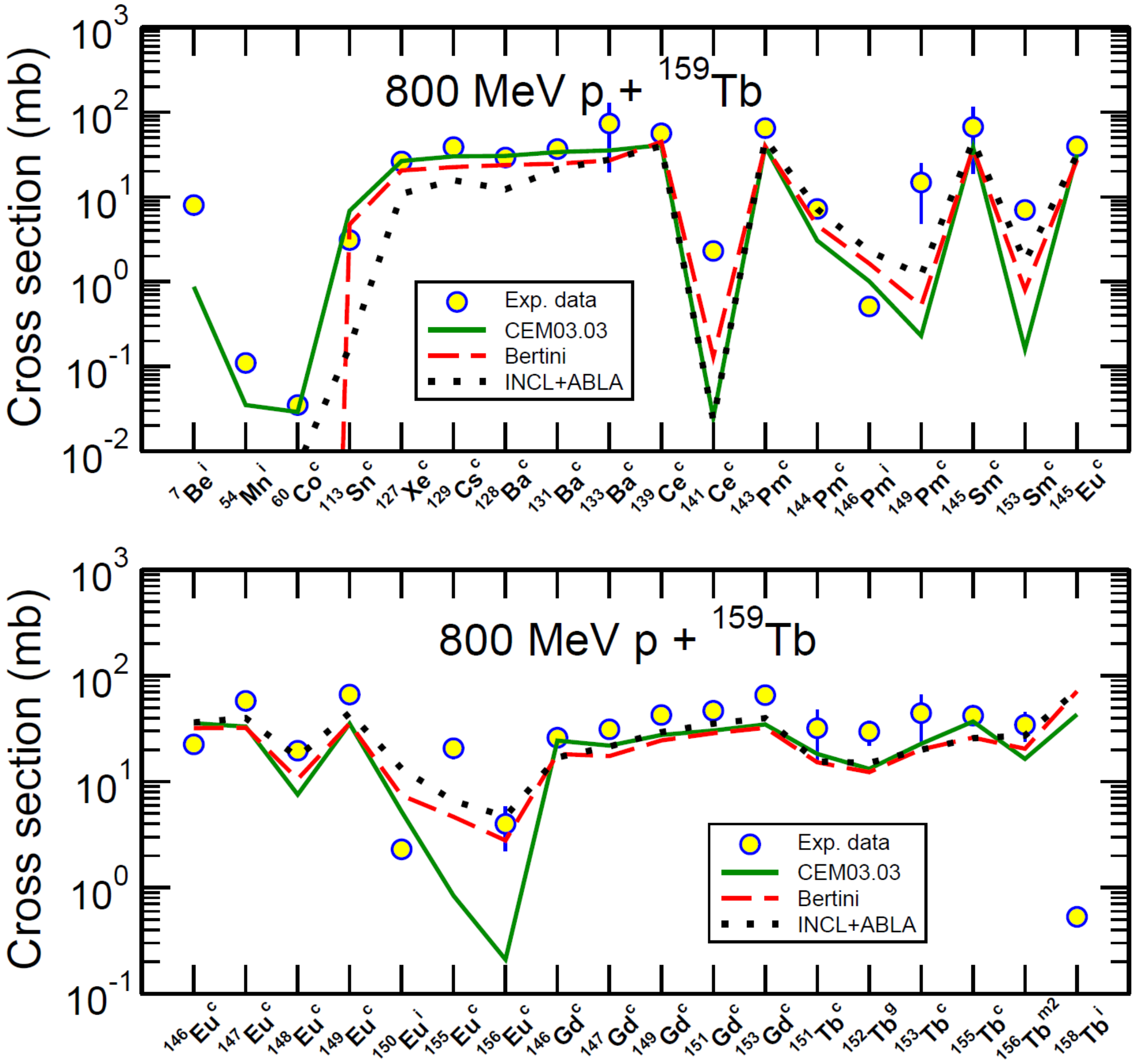

Fig. 3. Detailed comparison between all cross sections measured in the present work and those predicted by the CEM03.03, Bertini+MPM+Dresner+RAL, and INCL+ABLA event generators of MCNP6. The cumulative cross sections are labeled with a “c” and the independent cross sections, with an “i”.

## 5. Conclusions

Cross sections for 36 nuclides produced by the 800 MeV proton irradiation of terbium are measured for the first time and are compared with predictions by CEM03.03, Bertini+MPM+Dresner+RAL, and INCL+ABLA event generators of the MCNP6 transport code. Calculations by all event generators agree with the measured data in the mass region near A = 159 where spallation reactions dominate. CEM03.03 and INCL+ABLA predictions differ from one another in the region of fission fragments by an order of magnitude, with the CEM03.03 results much closer to the measured values. The RAL fission model used with the Bertini option of MCNP6 does not calculate at all fission of nuclei with atomic number Z < 70. Only one product, $^{7}Be$, could be measured in the fragmentation region. CEM03.03 predicts a cross section for $^{7}Be$ about nine times lower than the measured value, while INCL+ABLA predicts a still lower yield, and Bertini does not predict the formation of $^{7}$Be products from 800 MeV protons incident on terbium. In the region of fission fragments, CEM03.03 predicts the yield of $^{54}$Mn within a factor of 3 of the measured value and a yield of $^{60}$Co which is within experimental measurement uncertainty. The computational models studied here are expected to benefit from modifications to improve their predictive accuracy in light of these measured data.

## 6. Acknowledgements

We are grateful for technical assistance from LANL C-NR, C-IIAC, AOT-OPS, and LANSCE-WNR groups' staff and thank Dr. Arnold Sierk for useful discussions. This study was carried out under the auspices of the National Nuclear Security Administration of the U.S. Department of Energy at Los Alamos National Laboratory under Contract No. DE-AC52-06NA253996 with partial funding by the US DOE Office of Science via an award from The Isotope Development and Production for Research and Applications subprogram in the Office of Nuclear Physics.